
\magnification=\magstep1
\vsize=7.5in
\hsize=5.25in
\vskip 0.25in
\baselineskip=18pt
\tolerance=10000
\footline={\ifnum\pageno<2\hfil\else\hss\tenrm\folio\hss\fi}
\def\Dsl{\hbox{/\kern-.6000em\rm D}}

\def\Box{\vbox{\hrule\hbox{\vrule\kern3truept\vbox{\kern3truept\vbox{}
\kern3truept}\kern3truept\vrule}\hrule}}
\def\.{\hfill\break}

\def\hf{{1\over 2}}
\def\nth#1{{1\over #1}}

\def\ie{{\it i.e.}}

\def\apriori{{\it a priori}}

\def\cf{{\it c.f.}}

\def\etal{{\it et.al.}}

\def\p{{\bf p}}
\def\q{{\bf q}}

\def\today{\ifcase\month\or January\or February\or March\or
 April\or May\or June\or July\or August\or September\or
 October\or November\or December\fi \space\number\day, \number\year}

\newcount\refnumber
\gdef\clearrefnumber{\refnumber=0}
\def\refname#1{\xdef#1{\the\refnumber}}
\gdef\oneref#1{${}^{#1}$}

\gdef\ref#1{\advance\refnumber by1 ${}^{\the\refnumber}$ \refname{#1}}
\gdef\sref#1{\advance\refnumber by1 \refname{#1}}
\clearrefnumber

\newcount\fignumber
\gdef\clearfignumber{\fignumber=0}
\def\figname#1{\xdef#1{\the\fignumber}}

\gdef\fig#1{\advance\fignumber by1 Fig. [\the\fignumber] \figname{#1}}
\gdef\sfig#1{\advance\fignumber by1 \figname{#1}}
\clearfignumber

\newcount\tablenumber
\gdef\cleartablenumber{\tablenumber=0}
\def\tablename#1{\xdef#1{\the\tablenumber}}

\gdef\table#1{\advance\tablenumber by1 table. [\the\tablenumber]
\tablename{#1}}
\gdef\stable#1{\advance\tablenumber by1 \tablename{#1}}
\cleartablenumber

\newcount\notenumber

\def\notename#1{\xdef#1{\the\notenumber}}
\def\note#1#2{\advance\notenumber by1 \notename{#1}
\footnote{${}^{f\the\notenumber}$}{{\baselineskip=12pt #2}}}

\newcount\eqnumber
\gdef\cleareqnumber{\eqnumber=0}
\def\eqname#1{\xdef#1{\the\eqnumber}}
\gdef\eq{\the\eqnumber}

\gdef\eqn#1{\advance\eqnumber by1 \eqname{#1} }
\cleareqnumber

\tolerance=10000

\baselineskip=18pt
\magnification=1200
\vsize=23.0 true cm
\hsize=16.5 true cm
\rightline{McGill--91/24}
\rightline{July 1991}

\vskip 2 true cm
\centerline{ THE DAMPING OF ENERGETIC GLUONS AND QUARKS }
\centerline{ IN HIGH-TEMPERATURE QCD }
\vskip 1 true cm
\centerline { C.P. Burgess and A.L. Marini\footnote{${}^{*}$}{internet:
cliff@physics.mcgill.ca and alex@physics.mcgill.ca }\footnote{}{bitnet:
py30@mcgilla} }
\centerline{\sl Physics Department, McGill University, 3600 University St.}
\centerline{\sl Montr\'eal, Qu\'ebec, Canada, H3A 2T8. }
\vskip 1.5 true cm
\centerline{ABSTRACT}
\bigskip

\noindent
When a gluon or a quark is sent through the hot QCD plasma it can be absorbed
into the ambient heat bath and so can acquire an effective lifetime. At high
temperatures and for weak couplings the inverse lifetime, or damping rate, for
energetic quarks and transverse gluons, (those whose momenta satisfy $|\p|
\gg gT$) is given by $\gamma(\p) = c\; g^2 \log\left({1\over g}\right)\; T +
O(g^2T)$. We show that very simple arguments suffice both to fix the numerical
coefficient, $c$, in this expression and to show that the $O(g^2T)$
contribution is incalculable in perturbation theory without further
assumptions. For QCD with $N_c$ colours we find (expressed in terms of the
casimir invariants $C_a=N_c$ and $C_f=(N_c^2-1)/(2N_c)$): $c_g=+{C_a\over
4\pi}$ for gluons and $c_q=+{C_f\over 4\pi}$ for quarks. These numbers agree
with the more detailed calculations of Pisarski \etal\ but disagree with those
of Lebedev and Smilga. The simplicity of the calculation also permits a direct
verification of the gauge-invariance and physical sign of the result.

\vfill\eject

\pageno=2

The behaviour of nonabelian gauge theories at finite temperature is of
theoretical interest due to the surprisingly rich structure they exhibit even
within the perturbative regime of weak coupling and high temperatures. They may
also have phenomenological applications to the interpretation of the collisions
of heavy nuclei at high energies.

In recent years much attention has been directed towards
understanding the properties that gluons and quarks acquire as they propagate
through a quark-gluon plasma.\ref{\gluondamping}\  This has been at least
partially due to the sometimes contradictory and confusing results of the
earliest one-loop calculations. In these calculations the gluon damping
constant, $\gamma_g(\p)$, was found in the static ($|\p|\to 0$) limit to depend
on the gauge on which it was computed and, for some gauges, to be negative. If
taken seriously such a negative damping constant would indicate an instability
of the thermal state towards gluon emission. The source of the confusion with
these calculations\ref{\resolution}\ is that they neglect higher-loop
contributions that are of the same order in the gauge coupling constant, $g$,
as
are the terms that are kept. This is because in perturbation theory (in three
space dimensions) at finite temperatures successive terms in the loop expansion
need {\it not} be suppressed relative to fewer loops by additional factors of
the gauge coupling.

The failure of the loop expansion arises at finite temperatures because of the
occurence of severe infrared divergences.  These infrared divergences are more
troublesome than they are at zero temperature due to the singular behaviour of
the Bose-Einstein distribution function at low energies: $n(E) =
(e^{E/T}-1)^{-1}
\approx {T \over E} + \cdots$. This behaviour causes quantities to blow up like
a power of the infrared cutoff rather than simply logarithmically as they do at
zero temperature.\ref{\powercounting}\ Indeed, a simple power-counting argument
shows that for QCD at temperature $T$ a generic $\ell$-loop graph can
contribute an amount proportional to $(g^2 T/\lambda)^\ell$ relative to the
tree-level result. $\lambda$ in this expression is an infrared cutoff. Higher
loops clearly need not be suppressed once $\lambda$ is as low as $g^2T$.

In fact the loop expansion already fails even for an infrared cutoff as large
as $\lambda \approx gT$ since a subclass of diagrams can be even more infrared
singular than is indicated by the generic power-counting argument. The
dangerous
graphs are those such as the `ring' graphs within which multiple self-energy
insertions are made along a single internal line.\ref{\joesbook}\ In
reference\sref{\resummation} [\resummation] Braaten and Pisarski argue that
these last contributions may be resummed by dressing all `soft' lines---those
carrying momenta less than or of order $gT$---by the calculable contributions
of
`hard thermal loops'. In ref. [\resolution] they then compute the implications
for the particular case of the damping rate for static gluons, arguing that the
failure of perturbation theory is in this case completely cured by such a
resummation. This last conclusion is, however, disputed by Lebedev and
Smilga,\ref{\lebedev}\ who argue that other contributions beyond  the hard
thermal loops can contribute equally large effects to the static gluon damping
constant.

The purpose of this letter is to point out that for at least some physical
quantities the dominant part of the result for small $g$ can be identified
quite
simply without making use of the complete resummation formalism. We take by way
of illustration the damping constant, $\gamma(\p)$, for transverse gluons (and
 for quarks) but evaluated for momenta $|\p| \gg gT$ rather than in
the static limit. For these momenta the dominant contribution for small
coupling
to $\gamma(\p)$ is momentum independent and has the form $c
\;g^2\log\left({1\over g}\right) T + b\; g^2T +\cdots$. We show that the
coefficient $c$ can be computed by a simple, analytic calculation that only
involves wavelengths that are within the perturbative regime: $\lambda \gg g^2
T$. The same is not true for the coefficient $b$ which can receive
contributions
for $\lambda \approx g^2T$. We can in this instance therefore explicitly verify
that $\gamma(\p)$ is indeed gauge-independent and positive.

Our results, given in and immediately following eq. (12), may be compared to
more detailed calculations. For fermions they
agree with those of\sref{\heavyfermion} ref. [\heavyfermion] but disagree by a
factor of 3 with the real-time calculations of ref. [\lebedev]. For transverse
gluons we agree with the result of an as-yet-unpublished version of a full
resummation calculation,\ref{\private}\ but do not agree with the real-time
estimate of ref. [\lebedev].

We now turn to a description of our calculation. We focus first on the purely
gluonic theory since this is the case for which there is the most
simplification
over a full calculation.

The dispersion relation relating the energy and momentum for a relativistic
transverse gluon traversing the plasma may be determined by the position,
$E(\p)
= \omega(\p) -i\gamma(\p)$, in the complex energy plane of the zero of the
transverse part of the inverse gluon propagator. Using the usual one-parameter
family of covariant gauges for which the bare inverse Feynman propagator is:
\eqn{\covgauge}
$$ \left(G_{\rm bare}^{-1} \right)_{\mu\nu}(p) =  p^2 \eta_{\mu\nu} +
\left( \xi^{-1} - 1 \right) p_\mu p_\nu, \eqno(\eq)$$
and the full thermal inverse propagator is:
\eqn{\fullpropagator}
$$ \left(G_{\rm full}^{-1} \right)_{\mu\nu}(p) = \left(G_{\rm bare}^{-1}
\right)_{\mu\nu}(p) + \Pi_{\mu\nu}(E,\p), \eqno(\eq)$$
the dispersion relation for transverse gluons becomes:
\eqn{\massshell}
$$ E^2(\p) - \p^2 = \hf \left[ {\Pi^i}_i - {p^i p^j \Pi_{ij}\over
\p^2} \right]. \eqno(\eq)$$
Here $i$ and $j$ are to be summed over the three spatial directions,
$i,j=1,2,3$, in the plasma rest frame. Our task is to compute the leading
contributions to $\Pi_{\mu\nu}(E,\p)$ and to, in particular, identify the
origin
of the $g^2\log g$ terms.

In order to sort out the size of the various contributions to $\Pi_{\mu\nu}$
it is useful to divide the loop integrations according to whether or not they
involve only energies and momenta that are greater than some infrared cutoff,
$\lambda$. That is:
\eqn{\division}
$$\Pi_{\mu\nu}(E,\p) = \Pi^{\rm soft}_{\mu\nu}(E,\p,\lambda) + \Pi^{\rm
hard}_{\mu\nu}(E,\p,\lambda), \eqno(\eq)$$
in which all loop momenta in $\Pi^{\rm hard}_{\mu\nu}(E,\p,\lambda)$ are cut
off in the infrared at $\lambda$. Provided that $\lambda$ is chosen
sufficiently high $\Pi^{\rm hard}_{\mu\nu}$ is calculable within the loop
expansion. $\Pi^{\rm soft}_{\mu\nu}$ is not computable in this way and must be
obtained by other means. An important issue to be addressed is how much any
desired quantity depends on the largely undetermined $\Pi^{\rm soft}_{\mu\nu}$.

If $\lambda$ is chosen much greater than $gT$ then the lowest-order
contribution to $\Pi^{\rm hard}_{\mu\nu}$ arises at $O(g^2)$ due to
one-loop graphs. There are
four such one-loop graphs: one tadpole and one vacuum-polarization graph
having either an internal gluon, quark or ghost loop. For example, using the
Matsubara imaginary-time technique,\ref{\matsubara}\ the gluon
vacuum-polarization graph contributes an amount:
\eqn{\oneloop}
$$\eqalign{ \Pi_{\mu\nu}^{\rm hard}(E,\p,\lambda) &= {g^2 C_a T\over 2}
\sum_{n=-\infty}^\infty \int_{\q^2 > \lambda^2} {d^3\q \over (2\pi)^4} \;
V_{\mu\lambda\rho}(p,-p-q,q) \; G^{\lambda\kappa}(p+q) \cr
& \qquad \qquad \qquad \times \; G^{\rho\sigma}(q) \;
V_{\nu\sigma\kappa}(-p,-q,p+q).\cr} \eqno(\eq)$$
Here the gluon propagator is the inverse of eq. (\covgauge) and the three-point
gluon vertex is given by the zero-temperature expression:
$V_{\mu\lambda\rho}(p,q,k) = (p-q)_\rho \eta_{\mu\lambda} + (q-k)_\mu
\eta_{\rho\lambda} + (k-p)_\lambda \eta_{\mu\rho}$. The quadratic invariant in
the adjoint representation is $C_a=N_c$ ($=3$) for the gauge group $SU(N_c)$.
We work in Euclidean signature and the time component of every
four-vector is an integer times $2\pi T$. The summation is over the integer
corresponding to the loop momentum, $q$. The lower limit on the momentum
integration is meant as a reminder of the infrared cutoff.

If the infrared cutoff should instead be chosen to satisfy $g^2 T \ll \lambda
\ll gT$ then higher-loop graphs can contribute to the same order in $g$ as do
the one-loop graphs in which the loop momenta are of order $gT$. After
resumming
these higher-loop contributions the leading result for $\Pi^{\rm
hard}_{\mu\nu}$
is given by the same one-loop graphs as before, with the
proviso\oneref{\resummation}\ that each propagator (or vertex) is to be
replaced
with an `effective' resummed propagator (or vertex): $G_{\mu\nu}\to
G^*_{\mu\nu}$ and $V_{\mu\lambda\rho}\to V^*_{\mu\lambda\rho}$. Each of these
resummed quantities agrees (up to higher  powers of $g$) with its bare
counterpart unless all of the momenta entering the line (or vertex) in question
are themselves $O(gT)$. For such soft momenta, however, the resummed items are
relatively complicated functions of the ratios of the energies and momenta that
flow through that part of the graph. The main technical difficulty with working
with the effective propagators and vertices lies in manipulating this more
complicated form.

Some of this extra complication simplifies if the external momentum is itself
chosen to be much larger than $gT$, as is the case for when the dispersion
relation, $E(\p)$, is evaluated for hard momenta such as $|\p|\approx T$. The
reason for this simplification is that since the resummed and bare quantities
agree if any of the relevant momenta are hard, $|\p|\gg gT$, it is only
necessary to work with the resummed versions if {\it all} of the momenta
passing
through a particular line (or vertex) are soft. For hard external momenta many
of the internal lines and vertices must also carry hard momenta and so may be
represented by the usual bare Feynman rules. For example, in the
vacuum-polarization graphs both of the vertices and one propagator must
necessarily carry hard momenta if the external momentum is itself large in
comparison to $gT$. This leaves at most a single internal propagator to be
dressed.

Now comes the main point: the dominant part of $\Pi_{\mu\nu}$ for small
coupling---\ie\ the $g^2 \log g$ terms---can be determined with no knowledge of
$\Pi^{\rm soft}_{\mu\nu}$, and using virtually none of the complications of the
resummation formalism in $\Pi^{\rm hard}_{\mu\nu}$.

The principal observation is that all of the  terms in $\Pi_{\mu\nu}$ that are
proportional to $g^2 \log g$ are completely determined by {\it the infrared
divergent part} of $\Pi^{\rm hard}_{\mu\nu}$. To see how this works consider
the
lowest order contribution to the dispersion relation, $E(\p)$, for transverse
gluons with momenta $|\p|\sim T \gg gT$. This is determined from eq.
(\massshell) given the vacuum polarization, $\Pi_{\mu\nu}(E,\p)$ evaluated at
hard, on-shell momenta: $E=|\p| \sim T$. As is established in more detail
below,
the contribution of $\Pi^{\rm hard}_{\mu\nu}$ to the right-hand-side of eq.
(\massshell) diverges logarithmically with $\lambda$ for $g^2T \ll \lambda \ll
gT$:
\eqn{\generalform}
$$\eqalign{ F^{\rm hard}(T,|\p|,\lambda) &\equiv \hf
\left[ {(\Pi^{\rm hard})^i}_i - {p^i p^j \Pi^{\rm hard}_{ij}\over \p^2} \right]
\cr &= g^2 \left[ A \log\left( {\lambda \over \mu_{\rm hard}} \right) + B
+ O \left( {\lambda \over T} \right) \right] + O(g^3). \cr} \eqno(\eq)$$
In this expression $A$ and $B$ are purely functions of $|\p|$ and $T$ and
$\mu_{\rm hard}$ is a calculable energy scale of the high-frequency part of
the theory which turns out below to be $\mu_{\rm hard} \approx gT$.

In order to extract information about $F(T,|\p|) \equiv F^{\rm
hard}(T,|\p|,\lambda) + F^{\rm soft}(T,|\p|,\lambda)$ from eq. (\generalform)
it
is necessary to say something about the behaviour of $F^{\rm
soft}(T,|\p|,\lambda)$. The only property of $F^{\rm soft}(T,|\p|,\lambda)$
that
is required is that its $\lambda-$dependence must cancel that of $F^{\rm
hard}(T,|\p|,\lambda)$:
\eqn{\rengroup}
$$\eqalign{ \lambda {\partial F^{\rm soft}\over \partial \lambda} &\equiv -
\lambda {\partial F^{\rm hard}\over \partial \lambda}\cr
 &= g^2 \left[ -A + O\left( {\lambda \over T}\right) \right] + O(g^3).
\cr} \eqno(\eq)$$
This determines $F^{\rm soft}$ to have the form:
\eqn{\soft}
$$F^{\rm soft} = g^2 \left[ A \log\left( {\mu_{\rm soft} \over \lambda} \right)
+
C + O\left( {\lambda \over T}\right) \right] + O(g^3), \eqno(\eq)$$
in which $A$ is {\it the same function} as in eq. (\generalform). The constant
$\mu_{\rm soft}$ that appears within the logarithm in this equation is the
constant of integration that arises in passing from eq. (\rengroup) to eq.
(\soft). It has dimensions of mass and is chosen to be of order $g^2T$ since
this is the largest mass scale present in the soft part of the problem.

Adding the results of eqs. (\generalform) and (\soft) therefore gives:
\eqn{\theanswer}
$$\eqalign{ F &= - A g^2 \log\left({\mu_{\rm hard}\over \mu_{\rm soft}} \right)
+ g^2(B+C) + O\left({\lambda \over T} \right) + O(g^3) \cr
&= - A \; g^2 \log\left({1\over g}\right) + O(g^2),\cr} \eqno(\eq)$$
which determines the coefficient of the $g^2\log g$ term completely in terms of
the {\it calculable} coefficient $A$.

The next point is that since $A$ is determined by the infrared divergent part
of $F^{\rm hard}$, it is insensitive to most of the complications of the
resummation. To illustrate the simplicity with which $A$ may be determined we
now outline its calculation for the damping constant.

Inspection of the Feynman rules shows that the only potentially
infrared-divergent part of the leading contribution to $F^{\rm hard}$ comes
purely from the vacuum-polarization graph for $\Pi^{\rm hard}_{\mu\nu}$ in
which it is a gluon which circulates around the loop. Furthermore, even for
this graph an infrared divergence can arise only from the term for which the
integer
$n=q^0/(2\pi T)$ for the soft gluon line in eq. (\oneloop) vanishes,
and even then only if the external four-momentum is on shell: $E=|\p|$. Since
only $n=0$ contributes, it is sufficient to know the form for the resummed
propagator at zero frequency, where it reduces to:\oneref{\resummation}
\eqn{\resummedprop}
$$ \left[(G^*)^{-1}
\right]_{\mu\nu}(p) = \left(G_{\rm bare}^{-1} \right)_{\mu\nu}(p) + m^2
P_{\mu\nu}. \eqno(\eq)$$
Here $P_{\mu\nu}$ is the projection matrix onto the rest frame of the plasma,
and
so is given in this frame by the matrix diag$(0,1,1,1)$. $m$ denotes the lowest
order gluon mass, or plasma frequency, which is given in terms of the number of
quarks, $n_q$, by $m^2 = \nth{9} (gT)^2 \left( C_a + {n_q\over 2} \right)$.

Substituting this into eq. (\oneloop), evaluating on the lowest-order mass
shell, $E=|\p|\approx T$, and recognizing that at most one internal line can be
soft at a time (and so need be dressed) then gives an infrared-divergent
contribution:
\eqn{\irdivergence}
$$\eqalign{ \left. F^{\rm hard}(T,|\p|) \right|_{\rm div} &= -{g^2 C_a T m^2
|\p|\over 2 \pi^2} \int_\lambda^\infty dq \; {1\over q(q^2 + m^2)} \log\left({q
- 2|\p| + i\epsilon \over q + 2|\p|} \right) \cr
&= +{ig^2 C_a T |\p|\over 2\pi} \log \left({\lambda \over m}\right),\cr}
\eqno(\eq)$$ %
from which we read $A = (i C_a T |\p|)/(2\pi)$ and $\mu_{\rm soft}=m$. Notice
that the real part of
$F$ is infrared finite so only the lifetime acquires a $g^2 \log g$
contribution. Using this result in the mass-shell condition gives our main
result:
\eqn{\result}
$$\eqalign{ \gamma_g(\p) &= - {{\rm Im} \; F \over  2|\p|} \cr
&= + {g^2 C_a T \over 4\pi} \log \left({1\over g} \right) + O(g^2).\cr}
\eqno(\eq)$$

This entire argument when repeated for the quark self-energy similarly gives a
$g^2 \log g$ contribution to the branch of the fermion spectrum that survives
at large momenta. A simple calculation gives its coefficient as $\gamma_q(\p) =
{g^2 C_f T \over 4\pi} \log \left({1\over g} \right)+ O(g^2)$. $C_f$ denotes
the quadratic invariant in the fundamental representation:
$C_f=(N_c^2-1)/(2N_c)$ ($={4\over 3}$).

There are several features of this calculation that bear emphasis: ($i$) First,
as is required for a good approximation to a physical quantity, $\gamma$ is
independent of the gauge parameter $\xi$. All terms that depend
on the gauge-parameter contribute only an infrared-finite result. ($ii$) The
sign of $\gamma$ is positive, indicating stability.  ($iii$) Notice
that a determination of the subleading $O(g^2)$ contributions would require
knowledge of both of the coefficients $B$ of eq. (\generalform) and $C$ of eq.
(\soft). Although $B$ is calculable using the complete resummation formulation,
$C$ is not and can only at present be determined by making some assumptions
concerning the behaviour of the plasma in the low-frequency regime $\lambda
\approx g^2 T$. It follows that the coefficient $B$ need not \apriori\ by
itself
be gauge-independent (or positive). ($iv$) Also, since it is a logarithmic
infrared-divergence that is responsible for the logarithmic dependence on $g$,
its coefficient is insensitive to the details of how the cutoff is implemented.
Finally, ($v$) since the infrared-divergent term in $F^{\rm hard}$ is
explicitly
proportional to $m^2$ (\cf\ eq. (\irdivergence)) it only receives contributions
from the two-loop and higher graphs that serve to dress the soft propagator in
the gluon vacuum-polarization graph. It also follows that the imaginary part
arises only from the self-energy of internal lines which carry soft loop
momenta
$|\q| < gT$ since the mass $m$ may be taken to be zero for larger momenta. This
agrees with what is expected physically from unitarity given the constraints of
energy and momentum conservation in the plasma.

We conclude that for some quantities in which infrared divergences in the
perturbative expansion introduce a logarithmic dependence on the gauge
coupling,
$g$, it is possible to very simply identify the dominant contributions. This
simplicity allows a check on more complete and more involved calculations.

\noindent
ACKNOWLEDGEMENTS: The authors would like to acknowledge helpful conversations
with Charles Gale,
Joe Kapusta, Randy Kobes, Gabor Kunstatter and Rob Pisarski, as well as funding
by the Natural Sciences and Engineering Research Council of Canada and les
Fonds
pour la Formation de Chercheurs et l'Aide \`a la Recherche du Qu\'ebec.

\vfill\eject
\vskip 1.6truein
\noindent
{\bf REFERENCES}
\medskip

\def\anp#1#2#3{Ann. Phys. (NY) {\bf #1} (19#2) #3}

\def\npa#1#2#3{Nucl. Phys. {\bf A#1} (19#2) #3}
\def\npb#1#2#3{Nucl. Phys. {\bf B#1} (19#2) #3}
\def\plb#1#2#3{Phys. Lett. {\bf #1B} (19#2) #3}

\def\prd#1#2#3{Phys. Rev. {\bf D#1} (19#2) #3}

\def\prl#1#2#3{Phys. Rev. Lett. {\bf #1} (19#2) #3}
\def\ptp#1#2#3{Prog. Theor. Phys. {\bf #1} (19#2) #3}
\def\sjnp#1#2#3{Sov. J. Nucl. Phys. {\bf #1} (19#2) #3}
\def\rmp#1#2#3{Rev. Mod. Phys. {\bf #1} (19#2) #3}
\def\zpc#1#2#3{Zeit. Phys. {\bf C#1} (19#2) #3}

\item{[\gluondamping]}
O.K. Kalashnikov and V.V. Klimov, \sjnp{31}{80}{699};
D.J. Gross, R.D. Pisarski and L.G. Yaffe, \rmp{53}{81}{43};
U. Heinz, K. Kajantie and T. Toimela, \plb{183}{87}{96}, \anp{176}{87}{218};
K. Kajantie and J. Kapusta, \anp{160}{85}{477};
J.C. Parikh, P.J. Siemens and J.A. Lopez, unpublished;
T.H. Hansson and I. Zahed, \prl{58}{87}{2397}, \npb{292}{87}{725};
H.-Th. Elze, U. Heinz, K. Kajantie and T. Toimela, \zpc{37}{88}{305};
H.-Th. Elze, K. Kajantie and T. Toimela, \zpc{37}{88}{601};
R. Kobes and G. Kunstatter, \prl{61}{88}{392};
S. Nadkarni, \prl{61}{88}{396};
M.E. Carrington, T.H. Hansson, H. Yamagishi and I. Zahed, \anp{190}{89}{373};
J. Milana, \prd{39}{89}{2419};
G. Gatoff and J. Kapusta, \prd{41}{90}{611};
J. Kapusta and T. Tomeila, \prd{39}{89}{3197};
R. Kobes, G. Kunstatter and K.W. Mak, \plb{223}{89}{433};
S. Catani and E. d'Emilio, \plb{238}{90}{373}, \npb{}{}{};
K.A. James, \npb{}{}{}.

\item{[\resolution]}
R.D. Pisarski, \prl{63}{89}{1129}, \npa{525}{91}{175c};
E. Braaten and R.D. Pisarski, \prl{64}{90}{1338}.

\item{[\powercounting]}
S. Weinberg, in the proceedings of the International School of Subnuclear
Physics, {\it Ettore Majorana}, Erice, Sicily, Jul 23 - Aug 8, 1976
(QCD161:I65:1976).

\item{[\joesbook]}
For a discussion see, for example, J. Kapusta, {\it Finite Temperature Field
Theory}, (Cambridge University Press, 1985).

\item{[\resummation]}
E. Braaten and R.D. Pisarski, \prl{64}{90}{1338}, \npb{337}{90}{569}.

\item{[\lebedev]}
V.V. Lebedev and A.V. Smilga, University of Berne preprints BUTP-89/25 and
BUTP-90/38 (unpublished).

\item{[\heavyfermion]}
R.D. Pisarski, \prl{63}{89}{1129}; and E. Braaten and R.D. Pisarski,
unpublished.

\item{[\private]}
E. Braaten and R.D. Pisarski, unpublished.

\item{[\matsubara]}
T. Matsubara, \ptp{14}{55}{351}.

\vfill\eject

\bye